\documentclass[twocolumn]{aastex631}

\usepackage{amsmath}
\usepackage{mathrsfs}

\newcommand\bmm{{$\beta$-meteoroid }}
\newcommand\bms{{$\beta$-meteoroids }}
\newcommand\amsn{{$\alpha$-meteoroids}}
\newcommand\amm{{$\alpha$-meteoroid }}
\newcommand\ams{{$\alpha$-meteoroids }}
\newcommand\bmsn{{$\beta$-meteoroids}}

\begin{document}
\title{A Rotational Disruption Crisis for Zodiacal Dust} 

\correspondingauthor{Kedron Silsbee}
\email{kpsilsbee@utep.edu}

\author[0000-0003-1572-0505]{Kedron Silsbee}
\affil{University of Texas at El Paso, El Paso, TX, 79968}

\author[0000-0001-7449-4638]{Brandon S. Hensley}
\affil{Jet Propulsion Laboratory, California Institute of Technology, 4800 Oak Grove Drive, Pasadena, CA 91109, USA}

\author[0000-0003-2685-9801]{Jamey R. Szalay}
\affil{Princeton University, 171 Broadmead St., Princeton, NJ 08540}

\author[0000-0002-5667-9337]{Petr Pokorn{\'y}} 
\affil{Department of Physics, The Catholic University of America, Washington, DC 20064}

\author[0000-0001-6228-8634]{Jeong-Gyu Kim}
\affil{Quantum Universe Center, Korea Institute for Advanced Study, Hoegiro 85, Seoul 02455, Republic of Korea}

\begin{abstract}
A systematic torque from anisotropic radiation can rapidly spin up irregular grains to the point of breakup. We apply the standard theory of rotational disruption from radiative torques to solar system grains, finding that grains with radii $\sim$0.03\,--3\,$\mu$m at 1\,au from the Sun are spun to the point of breakup on timescales $\lesssim1$\,yr even when assuming them to have an unrealistically high tensile strength of pure meteoritic iron. Such a rapid disruption timescale is incompatible with both the abundance of micron-sized grains detected in the inner solar system and with the low production rate of $\beta$ meteoroids. We suggest the possibility that zodiacal grains have a strong propensity to attain rotational equilibrium at low angular velocity (a so-called low-$J$ attractor) and that the efficacy of rotational disruption in the Solar System---and likely elsewhere---has been greatly overestimated.
\end{abstract}

\section{Introduction}

Solar radiation has a strong influence on the dynamics of small bodies in the Solar System. Poynting-Robertson drag causes the orbits of small particles to decay \citep{Robertson:1937, Burns:1979}. Over long time periods, variations in the absorption and emission of solar photons due to diurnal and seasonal cycles alter the orbital radius of bodies around the Sun in a process known as the Yarkovsky effect \citep{Opik:1951}. In addition to modifying the body's orbit, solar radiation can also change its rotation rate and orientation in what is termed the Yarkovsky-O'Keefe-Radzievskii-Paddack (YORP) effect \citep{Rubincam:2000}. While the YORP effect is typically considered in the limit of macroscopic irregularities in large bodies, an analogous effect operates when there are anisotropies in the body on scales comparable to the wavelength of the illuminating radiation \citep{Silsbee16}. In this work, we study the consequences of this effect on the rotation rate of small ($\ll 10$\,cm) bodies.

If a body is spun up to sufficiently high angular speeds, it will break apart. The first consideration of this effect in the Solar System posited that albedo variations on the surfaces of particles would cause solar radiation to induce a systematic torque large enough to spin small bodies to breakup \citep{Radzievskii:1952}. Later, it was noted that small bodies with an irregular shape exposed to solar radiation would undergo a ``windmill effect,'' analogous to how leaves rotate when falling to Earth or small stones spin when dropped into a swimming pool, that over time would lead to ``rotational bursting'' \citep{Paddack69, Paddack:1973, Paddack:1975}. \citet{O'Keefe:1976} referred to this destruction mechanism as the Radzievskii-Paddack effect. YORP-induced rotational disruption has been shown to significantly alter the size-distribution of asteroids less than several kilometers in size \citep{Bottke:2006, Jacobson14}.

Beyond the Solar System, radiative torques are thought to be critical to the dynamics of dust grains in the interstellar medium (ISM). Shortly following the discovery that starlight is polarized coherently across many degrees on the sky \citep{Hall:1949, Hiltner:1949}, it was realized that interstellar grains must be aspherical and systematically aligned, polarizing starlight via dichroic absorption \citep{Davis:1951}. In order for grain alignment not to be randomized by thermal fluctuations, grains must rotate suprathermally \citep{Purcell:1979}. While \citet{Purcell:1979} proposed a systematic torque from H$_2$ formation on grain surfaces, \citet{Draine:1996} demonstrated that radiative torques from anisotropies in the ambient interstellar radiation field operate much faster. Because radiative torques are effective only for grains of at least comparable size to the wavelength of the illuminating radiation, they naturally explain the fact that optical extinction is more highly polarized than UV extinction, which arises from smaller grains \citep[see discussion on the size distribution of aligned grains in][]{Kim:1995}.

Given the relatively short timescales on which radiative torques spin up grains in the ISM, it was realized that sub-micron-sized grains would also be subject to rotational disruption if their angular velocity becomes sufficiently large \citep{Silsbee16, Hoang19}. Referred to as the radiative torque disruption (RATD) mechanism in much recent literature \citep[][and subsequent work]{Hoang19}, this effect has now been invoked in a wide range of astrophysical contexts: in the vicinity of massive stars and supernovae \citep{Hoang19}, cometary dust \citep{Herranen20}, protoplanetary disks \citep{Tung20}, molecular clouds \citep{Hoang21MC}, the solar F-corona \citep{Hoang21}, and exoplanet atmospheres \citep{Hoang23}.

Despite the potentially far-reaching applications of rotational disruption, substantial uncertainties remain in the efficacy of the mechanism. Much of the recent theoretical work has assumed grains can be driven to steady-state with a rapid angular velocity and large angular momentum $J$, a so-called high-$J$ attractor. However, bodies can also reach a stable equilibrium at much slower rotation rates, i.e., low-$J$ attractors. It has been argued that $J$-randomizing processes, such as collisions with gas atoms, can effectively move bodies from low-$J$ attractors to high-$J$ attractors while having little effect on bodies already in a high-$J$ state \citep{Hoang16}. Not all bodies possess a high-$J$ attractor state \citep{Draine:1996, Lazarian:2007}, although it was shown by \citet{Hoang16} that the presence of superparamagnetic inclusions in the grain increases the fraction of grains containing such states to near 100\%. Given the large uncertainties in the material properties of grains and in the microphysics of rotational spinup, observational tests are required.

The zodiacal cloud is an ideal laboratory for testing rotational disruption by virtue of having both a strongly anisotropic radiation field and a large complement of direct and indirect measurements of the size and spatial distribution of small particles. In this work, we argue that straightforward application of current rotational disruption theory to solar system dust implies destruction rates far in excess of what is allowed by observations. A reconsideration of the physics of radiative spinup, and in particular of the propensity of grains to attain high-$J$ attractor states, is warranted.

The paper is organized as follows: Section~\ref{sec:theory} is a self-contained analysis of the rotational dynamics of grains in the Solar System; Section~\ref{sec:results} compares our disruption-time calculations with other relevant timescales for solar system grains; Section~\ref{sec:observations} tests the predictions of so short a rotational disruption timescale against a suite of observations, finding these timescales implausible; Section~\ref{sec:discussion} discusses possible resolutions; and Section~\ref{sec:summary} concludes with a summary.

\section{Calculation of rotational disruption in the solar system} \label{sec:theory}

\subsection{Grain Properties}
In this section, we investigate the rotational dynamics of solar system grains as a function of their size. For simplicity, we assume that all grains are spheres of radius $a$ having the dielectric function of ``astrodust'' developed for modeling dust in the diffuse ISM \citep{Draine:2021}. We use the dielectric function of the standard astrodust model that assumes that grains are oblate spheroids with axial ratio 1.4:1 and a porosity (defined as fraction of the grain volume filled by vacuum) of 0.2 \citep{Hensley:2023}, although we do not invoke asphericity or porosity in this work. We adopt the astrodust mass density in this model of $\rho = 2.74$\,g\,cm$^{-3}$. Solar System grains are unlikely to have the same composition as interstellar grains, but we do not anticipate that our results depend sensitively on the adopted optical properties or density. Finally, we assume that grains have a tensile strength of meteoritic iron, $S_{\rm max} = 5\times10^9$\,dyn\,cm$^{-2}$ \citep{Ahles:2021}, to be maximally conservative on the ability of radiative torques to disrupt grains.   

Given a grain size and our assumed dielectric function, we use Mie theory to compute the wavelength-dependent absorption, scattering, and extinction cross sections $C_{\rm abs}$, $C_{\rm sca}$, and $C_{\rm ext}$. These are often expressed as efficiency factors $Q$ relative to the geometrical cross section of the grain defined as $Q_{\rm X} \equiv C_{\rm X}/\pi a^2$, where X is one of absorption, scattering, or extinction.

The exception to this simplified grain model is the computation of radiative torques in Section~\ref{subsec:rats}. Spherical grains have no  helicity and so do not experience a radiative torque, unlike realistic irregular grains. Rather than employ a more complicated grain geometry throughout, we instead assume that the radiative torque efficiency is a known function of grain size irrespective of our assumption on grain shape [Equation~\eqref{eq:Qgamma}]. While this raises a technical inconsistency, we do not anticipate that our analysis of other aspects of the grain physics (e.g., heating, absorption, emission) would change appreciably for irregular shapes versus spheres whereas the computations would become significantly more complex.

\subsection{Grain Temperature}
The energy density in radiation of wavelength $\lambda$ at a distance $r \gg R_\odot$ from the Sun is

\begin{equation}
    u_\lambda\left(r\right) = \frac{\pi}{c} \left(\frac{R_\odot}{r}\right)^2 B_\lambda(T_\odot)
    \,,
\end{equation}
where $B_\lambda\left(T\right)$ is the Planck function, $c$ is the speed of light, $T_\odot = 5777$\,K, and $R_\odot = 6.957\times10^{10}$\,cm. The power absorbed by a grain of radius $a$ located at a distance $r$ is then

\begin{equation}
    P_{\rm abs}(a, r) = \pi a^2 c \int_0^\infty u_\lambda\left(r\right) Q_{\rm abs}(a, \lambda) d \lambda
    \,.
\end{equation}
If the grain is heated to a steady-state temperature $T_{\rm gr}$, it emits power

\begin{equation}
    P_{\rm em}(a, T_{\rm gr}) = 4 \pi^2 a^2 \int_0^\infty B_\lambda(T_{\rm gr}) Q_{\rm abs}(a, \lambda) d \lambda
    \,.
\end{equation}
We find $T_{\rm gr}$ by numerically solving

\begin{equation}
    P_{\rm abs}(a, r) = P_{\rm em}(a, T_{\rm gr})
    \,.
\end{equation}

\subsection{Grain Rotational Dynamics} \label{sec:rotation}

\subsubsection{Overview}
The steady-state rotation rate of a Solar System grain is set by the balance of radiative torques acting to spin it up and torques acting to damp its rotation. As we argue in Section~\ref{subsec:ir_torque}, damping is dominated by the emission of infrared photons.  Thus,

\begin{equation} \label{eq:omega_ode}
    I_{\rm gr} \dot \omega = \Gamma_{\rm rad} + \Gamma_{\rm IR}
    \,,
\end{equation}
where $I_{\rm gr}$ is the moment of inertia of the grain, $\omega$ is the angular frequency, $\Gamma_{\rm rad}$ is the radiative torque, and $\Gamma_{\rm IR}$ is the damping torque from infrared emission. Note that in this convention, $\Gamma_{\rm IR}$ is negative. Since we assume spherical grains,

\begin{equation}
    I_{\rm gr} = \frac{8 \pi}{15} \rho a^5
    \,.
\end{equation}

Calculation of $\Gamma_{\rm IR}$ and $\Gamma_{\rm rad}$ is discussed in the following sections.

\subsubsection{Radiative Torques} \label{subsec:rats}
A grain in an anisotropic radiation field experiences a torque

\begin{equation}
    \Gamma_{\rm rad} = \frac{1}{2} a^2 \int_0^\infty  \lambda u_\lambda \gamma Q_\Gamma(\lambda) d \lambda
    \,,
    \label{eq:radiativeTorque}
\end{equation}
where $\gamma$ is a scalar parameter describing the anisotropy of the radiation field ($\gamma = 1$ for grains illuminated by the Sun) and $Q_\Gamma$ is the radiative torque efficiency factor \citep{Draine:1996}. $Q_\Gamma$ depends on the optical properties of the grain as well as the direction and polarization of the incident radiation relative to the grain body.

We follow previous studies \citep[e.g.,][]{Hoang19} and adopt the empirical scaling of $Q_\Gamma$ with grain size found by \citet{Lazarian:2007}:

\begin{equation}
    Q_\Gamma\left(a,\lambda\right) = 
    \begin{cases}
        0.4 \left(\frac{\lambda}{1.8a}\right)^{-3}\,,\quad &\lambda \geq 1.8a \\
        0.4\,,\quad &\lambda < 1.8a
    \end{cases}
    \,,
    \label{eq:Qgamma}
\end{equation}
Recently \citet{Jaeger:2024} calculated the radiative torque efficiency for ballistic aggregate dust models. While their results are broadly similar to those of \citet{Lazarian:2007}, they find a substantially reduced torque cross-section for intermediate value of $\lambda/a$.  Using their Equation~(13) with the parameters taken for the BA geometry for carbonaceous grains (their Table~A1), we find that the radiative torque cross-section averaged over the solar spectrum is smaller by factors of [0.38, 0.78, 0.74] than the prescription in \citet{Lazarian:2007} [Equation~\eqref{eq:Qgamma}] for grains with sizes of [0.1, 1, 10]\,$\mu$m, respectively.

\subsubsection{Rotational damping} \label{subsec:ir_torque}

\citet{Draine98} identified several mechanisms by which rotational motion of dust grains is damped: collisions with neutral particles, collisions with ions, plasma drag, emission of infrared photons, and electric dipole emission.  For each process $p$, they calculate a damping rate $F_p$ normalized to the damping of an electrically neutral grain due to collisions with neutral hydrogen atoms of density $n_{\rm H}$ that stick temporarily to the surface.  

To estimate the relative contribution of these damping mechanisms, we take as a reference system a 1\,$\mu$m grain, with a dipole moment of 0.4 Debye$\times \sqrt{N_A}$ where $N_A$ is the number of atoms \citep{Draine98}.  We assume that this is  embedded in a fully ionized $10^5$\,K hydrogen plasma with $n_{\rm H} = 10$ cm$^{-3}$, and a radiation field appropriate for a distance 1\,au from the Sun.  Using Equation~(30) from \citet{Draine98}, we calculate $F_{\rm IR}$ (the normalized rate of damping due to emission of infrared photons) equal to 500.  Equation~(25) from \citet{Draine98} shows that the normalized damping rate due to plasma drag is of order $10^{-5}$.  Similarly, damping due to electric dipole emission is subdominant by approximately 16 orders of magnitude for a micron-sized grain spinning at break-up speed.  Dust grains in the inner solar system are expected to be somewhat positively charged \citep{Kimura98}, and therefore the normalized damping rate due to collisions with positive ions should be less than unity (as the collision rate is suppressed by the Coulomb barrier).  Therefore, we conclude that damping due to emission of infrared photons is dominant by a factor greater than the calculated $F_{\rm IR} = 500$, and we include only this damping process.

As a rotating grain emits radiation, it loses angular momentum \citep{Purcell:1971}. In the limit $a/\lambda \ll 1$ (the ``electric dipole limit''), \citet{AHD09} showed that the rotational damping is given by 

\begin{equation}
    \Gamma_{\rm IR}^{\rm ed} = -\frac{2 \omega}{\pi} \int_0^\infty \frac{F_\lambda}{\nu^2} d \lambda,
    \label{eq:gammadamp}
\end{equation}
where

\begin{equation} \label{eq:Flambda}
    F_\lambda = \pi a^2 Q_{\rm abs}(a, \lambda) B_\lambda(\lambda, T_{\rm gr})
\end{equation}
is the power emitted by the grain per steradian per wavelength interval $d\lambda$.

The effects of rotational damping by photon emission were considered in the large particle (``LP'') limit by \citet{Jones:1990}, who found that a grain emitting energy at a rate $\dot{E}$ loses angular momentum $L$ at a rate

\begin{equation}
    {\dot L_{\rm IR}}^{\rm LP} = \frac{2a^2\omega}{3c^2}\dot{E}
    \,.
\end{equation}
Therefore, if the grain radiates total power per steradian $F_\lambda$, the torque on the grain is

\begin{equation} \label{eq:gammaIR_large}
    \Gamma_{\rm IR}^{\rm LP} =  -\int_0^\infty d \lambda \frac{8\pi a^2F_\lambda\omega}{3c^2}
    \,.
\end{equation}

While the above expressions yield $\Gamma_{\rm IR}$ in the limits of $a \ll \lambda$ and $a \gg \lambda$, we are unaware of an expression valid in the intermediate regime of $a\sim\lambda$. We instead employ the simple interpolation

\begin{equation}
    \Gamma_{\rm IR} \approx \Gamma_{\rm IR}^{\rm ed} + \Gamma_{\rm IR}^{\rm LP} = -\int_0^\infty d \lambda \frac{F_\lambda \omega}{c^2} \left(\frac{2 \lambda^2}{\pi} + \frac{8 \pi a^2}{3}\right)
    \,,
\end{equation}
which has the correct asymptotic behavior in the limits of large and small $a/\lambda$. 

Accounting for stronger damping in the large-grain regime, rather than using Equation~\eqref{eq:gammadamp} for all grains, is the only conceptual difference between our model for rotational disruption and that in \citet{Hoang19}. The difference is important for grains larger than approximately $5 ({\rm r/au})^{0.5}\,\mu$m.

\subsubsection{Equilibrium Rotation Rate}
The equilibrium angular velocity $\omega_{\rm eq}$ can be determined by first noting that both $\Gamma_{\rm rad}$ and $\xi \equiv \Gamma_{\rm IR}/\omega$ are independent of $\omega$. Then, solving Equation~\eqref{eq:omega_ode}, a grain spun up from rest at time $t = 0$ has

\begin{equation} \label{eq:omega_t}
    \omega\left(t\right) = \omega_{\rm eq}\left[1-{\rm exp}\left(-t/\tau_{\rm spinup}\right)\right]
    \,,
\end{equation}
where

\begin{equation}
    \omega_{\rm eq} = -\frac{\Gamma_{\rm rad}}{\xi},
\end{equation}
and

\begin{equation}
    \tau_{\rm spinup} = -\frac{I_{\rm gr}}{\xi}
    \,.
\end{equation}
Thus, the rotation rate asymptotes to $\omega_{\rm eq}$ as $t \rightarrow \infty$.

\subsection{Rotational Disruption}

\subsubsection{The Rotational Disruption Mechanism}
At some $\omega_{\rm crit}$, the grain angular velocity becomes so rapid that the grain breaks apart. Following \citet{Silsbee16}, we take $\omega_{\rm crit}$ to be the value of $\omega$ at which the centrifugal stress pulling two hemispheres of a grain apart is equal to the yield stress of the material $S_{\rm max}$ multiplied by the cross-sectional area:

\begin{equation} \label{eq:omega_crit}
\omega_{\rm crit} = \frac{2}{a} \sqrt{\frac{S_{\rm max}}{\rho}}
\,.
\end{equation}
Since the stress is not uniform across a plane separating the two hemispheres, the stress will locally exceed $S_{\rm max}$ before $\omega$ reaches $\omega_{\rm crit}$, thus breaking the grain at that point. Therefore, Equation~\eqref{eq:omega_crit} provides an upper bound to the critical frequency.  

The time $\tau_{\rm disrupt}$ to spin up a grain from $\omega = 0$ to $\omega_{\rm crit}$ can then be computed from Equation~\eqref{eq:omega_t}:

\begin{equation} \label{eq:disruptionTime}
    \tau_{\rm disrupt} = \tau_{\rm spinup} \ln\left(\frac{\omega_{\rm eq}}{\omega_{\rm eq} - \omega_{\rm crit}}\right)
    \,.
\end{equation}

\subsubsection{Rotational Disruption Size Limits}
\begin{figure}
\centering
\includegraphics[width = \columnwidth]{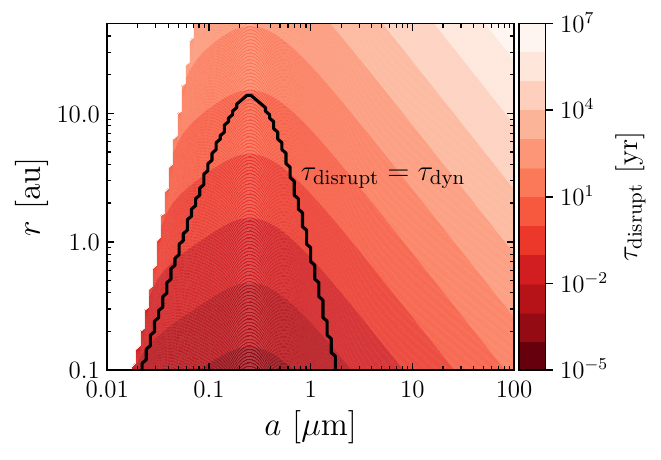}
\caption{Disruption timescale given by Equation~\eqref{eq:disruptionTime} as a function of grain size and heliocentric distance.  The black curve corresponds to the location in this space where the disruption time is equal to the dynamical time (see Equation \eqref{eq:tdyn}).  The white space on the left side of the figure corresponds to the region in which $\omega_{\rm eq} < \omega_{\rm crit}$, so the grain is not disrupted.}
\label{tauColorPlot}
\end{figure}

Figure~\ref{tauColorPlot} shows the disruption time in the space of grain size and orbital radius.  Grains between approximately 0.05 and 1~$\mu$m disrupt within $\tau_{\rm dyn}$ at 1~au, where $\tau_{\rm dyn}$ is the nominal dynamical time defined as the orbital radius divided by the orbital velocity, assuming a circular orbit:
\begin{equation}
\label{eq:tdyn}
\tau_{\rm dyn} = \frac{1{\rm yr}}{2\pi} \left(\frac{{\rm r}}{\rm au}\right)^{3/2}\,.
\end{equation}
This is the range of grain sizes for which one would expect orbital motion to have no appreciable effect on the rotational dynamics, and thus the survival of such grains puts the most stringent constraints on the rotational destruction mechanism described in \citet{Hoang19}.

Figure~\ref{tauColorPlot} suggests that rotational disruption is most effective for grains with sizes comparable to the mean wavelength of the solar spectrum.  Grains with $a \ll \lambda$ are not effectively spun up by radiative torques [Equation~\eqref{eq:Qgamma}] and so never reach $\omega_{\rm crit}$.  We find a lower size cut-off ranging from 20-70~nm at distances from 0.1 to 50~au from the Sun.   Grains with $a \gg \lambda$ undergo effective damping from infrared emission [Equation~\eqref{eq:gammaIR_large}] and so likewise never reach $\omega_{\rm crit}$.  Our calculations predict an upper cutoff of 3.5\,cm that is essentially independent of heliocentric distance from 0.1--50~au.  

Calculation of this upper cutoff is uncertain for two reasons. First, Equation~\eqref{eq:Qgamma} was derived by  \citet{Lazarian:2007} in the limit  $\lambda/a \geq 0.1$. If the radiative torque arises from features on the grain surface large compared to the wavelength, then for a fixed grain shape the torque should scale with $a^3$ for $\lambda/a \ll 1$ rather than the $a^2$ implied by Equations~\eqref{eq:radiativeTorque} and \eqref{eq:Qgamma}. If this argument is correct, then Equation~\eqref{eq:Qgamma} underestimates the radiative torque on large grains. Second, if the timescale for spinup is longer than the orbital time, then the excitation torque may partially cancel out due to averaging over the direction of the radiation field anisotropy. This effect would result in Equation~\eqref{eq:Qgamma} overestimating the net torque over the grain lifetime. The effects of orbital averaging are discussed further in Section~\ref{sec:discussion} and in Appendix~\ref{app:orbitAveraging}. As the main conclusions of this work are for small grains with spinup time short compared to their orbital time, they are not affected by these uncertainties.

\section{Comparison With Other Mechanisms Of Dust Removal}
\label{sec:results}
 \begin{figure*}[t!]
\centering
\includegraphics[width = \textwidth]{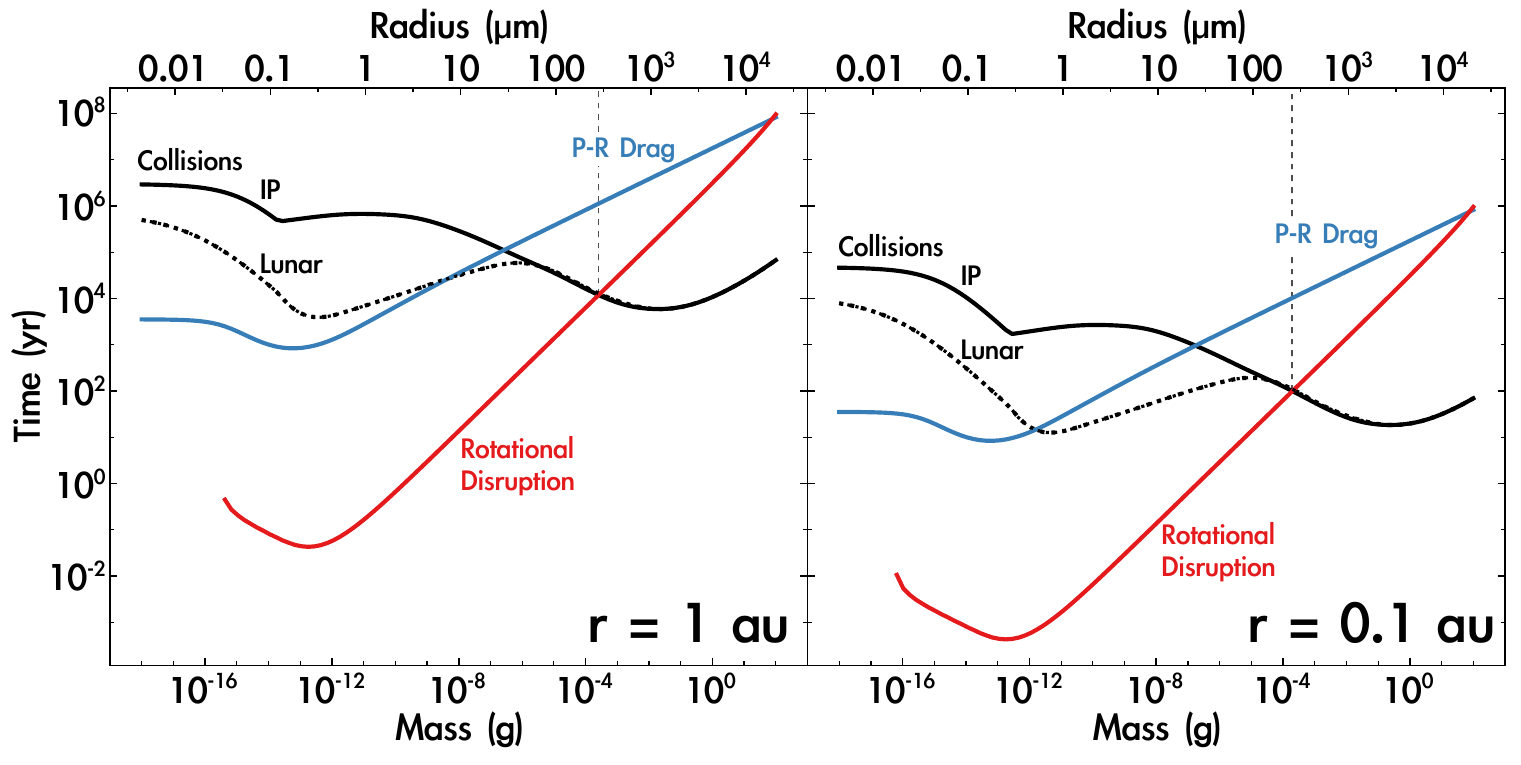}
\caption{Comparison of the timescales for grain removal by collisions \citep[assuming two different grain size distributions --- ``IP" = ``Interplanetary"; see][]{Gruen85}, Poynting-Robertson drag, and rotational disruption.  The two plots are made for different orbital radii as labeled on the panels.}
 \label{GruenPlot}
\end{figure*}

Here we compare the disruption timescale with two other relevant timescales for removal of Solar System grains: collisions with other grains and Poynting-Robertson drag. We evaluate the collisional lifetime of grains following the model in \citet{Gruen85}. We define the Poynting-Robertson drag timescale as

\begin{equation}
\tau_{\rm PR} = \frac{r}{\dot r} = \frac{r^2c}{2GM\beta},
\end{equation}
where $\beta$ is the ratio of radiation pressure force to gravitational force, given by 
\begin{equation}
    \beta = \frac{3\langle Q_{\rm pr}\rangle L_\odot}{16 \pi GM_\odot ac\rho}.
\end{equation}
Here, $\langle Q_{\rm pr}\rangle$ is the radiation pressure efficiency factor averaged over the solar spectrum.
This is a factor of two longer than the timescale reported in \citet{Gruen85}, which was based on the time required for the grain to spiral into the Sun, $r/(2\dot r)$.

The timescales for these processes are compared in Figure~\ref{GruenPlot}. At orbital radii of both $r = 1$\,au (left panel) and $r = 0.1$\,au (right panel), for particles with sizes between 30\,nm and 300\,$\mu$m, the rotational disruption timescale is shortest, generally by a few orders of magnitude.

Therefore, solar system grains with radii between roughly 30\,nm and a few microns have rotational disruption timescales much shorter than other removal timescales and shorter than the orbital timescale. If rotational disruption operates as described by this formalism, few grains in this size range could persist over a wide range of orbital radii even assuming unrealistically large tensile strengths. We next confront this prediction with observations.

\section{Comparison with Observations} \label{sec:observations}

\subsection{Overview}
In situ measurements of interplanetary dust grains in the micron-sized range have been performed for more than 50 years. Dedicated dust instruments that measured interplanetary grains were flown on Pioneer~8 and 9 \citep{berg:73b}, HEOS-2 \citep{hoffmann:75a, hoffmann:75b}, Helios~1 and 2 \citep{grun:80a, altobelli:06a, kruger:20a}, Ulysses \citep{grun:92b, wehry:99a, wehry:04a}, Galileo\citep{grun:92a,grun:92d}, Cassini \citep{srama:04a,kempf:04a}, and New Horizons \citep{poppe:10a, doner:24a}. 

Additionally, several space missions detected micron- and/or submicron-sized dust impacts using electric field-based observations of high-amplitude brief voltage spikes caused by dust impact-generated plasma: Voyager~2 \citep{gurnett:83a}, Vega \citep{laakso:89a}, DS-1 \citep{tsurutani:03a,tsurutani:04a}, Wind \citep{malaspina:14a,kellogg:16a}, MAVEN \citep{andersson:15a}, STEREO \citep{zaslavsky:12a}, MMS \citep{vaverka:18a,vaverka:19a}, Solar Orbiter \citep{zaslavsky:21a}, and Parker Solar Probe \citep{szalay:20a, malaspina:23a}. 

Grains in the zodiacal cloud can be dynamically grouped into two distinct populations: 1) \amsn, grains bound to the Sun's gravity in elliptic orbits, and 2) \bmsn, grains which experience enough outward solar radiation pressure that they are on unbound, hyperbolic trajectories. We note there are two definitions of \ams in the literature \citep{sommer:23a}: a) a dynamical subset of gravitationally bound grains with very large eccentricities and b) all grains gravitationally bound to the Sun. We use the latter, broad definition for \amsn. 

\begin{figure*}
\centering
\includegraphics[width = 0.98\textwidth]{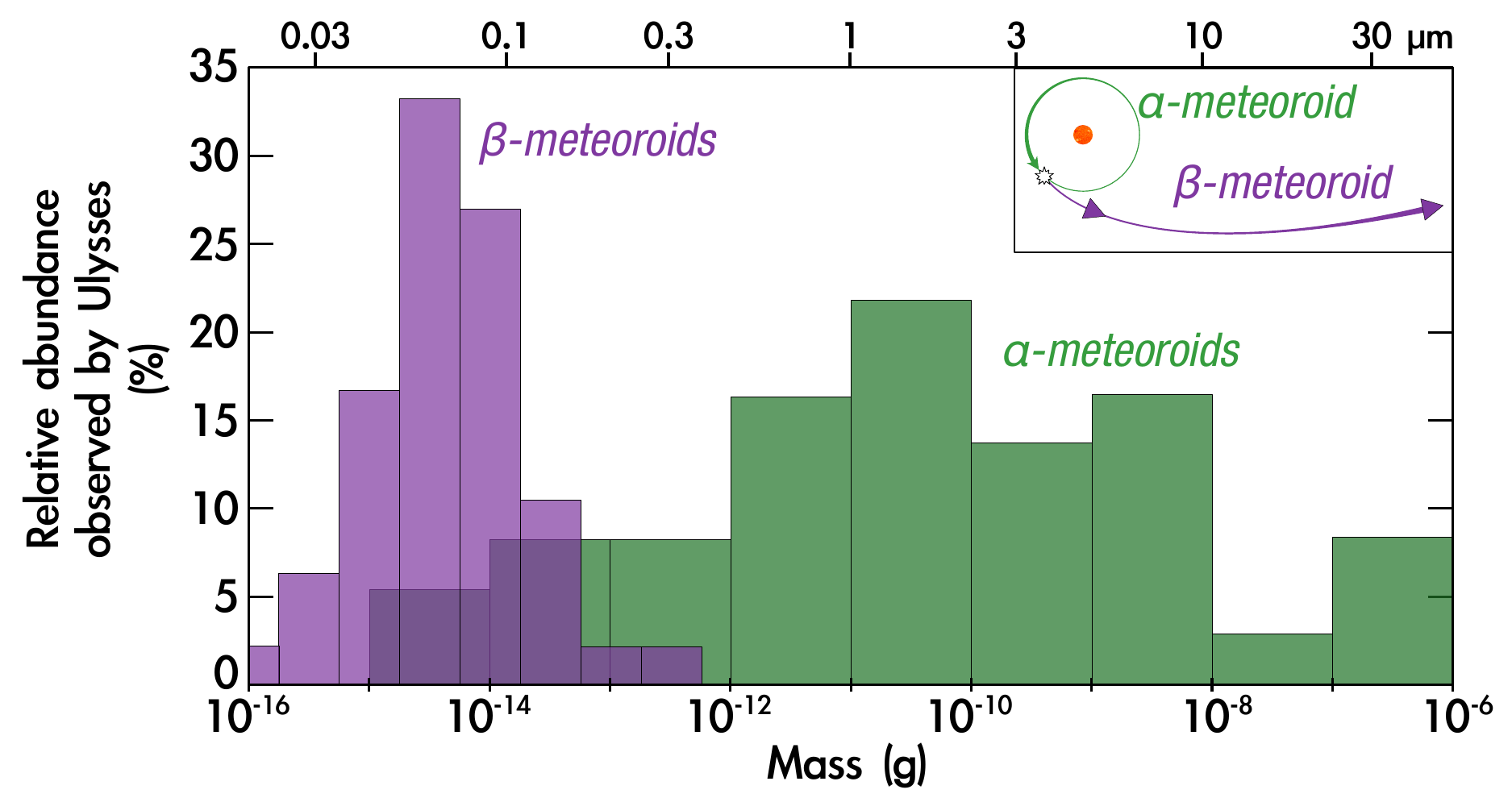}
\caption{Mass distribution observed by Ulysses for \ams near the equatorial plane in green \citep{grun:97a} and \bms observed throughout the mission in purple \citep{wehry:99a}. The distribution of \ams also has a minor contribution from interstellar grains in the lowest mass range \citep{grun:97a}. The top axis shows the mass converted to radius assuming ``old cometary'' grains \citep{wilck:96a}. All \amm masses shown here would be rotationally disrupted on very short timescales with the current assumptions on rotational distruption physics.}
 \label{fig:meteoroids}
\end{figure*}

The mass distributions of $\alpha$- and $\beta$-meteoroids inferred from Ulysses observations at heliocentric distances of 1.3 to 5\,au \citep{grun:97a, Wehry:1999} are summarized in Figure~\ref{fig:meteoroids}. Both populations are readily observed despite having sizes that are most susceptible to rotational disruption. Additionally, bound \ams with radii $\sim$1\,$\mu$m dominate the impact rates experienced by PSP as near as 0.1\,au from the Sun \citep{szalay:24a}. Grains with such small radii are not susceptible to collisional break-ups during their lifetime due to their short Poynting-Robertson drag timescales and long collisional lifetimes \citep{Pokorny:2024}.  There may be some contamination of the $\alpha$-meteoroid signal from interstellar dust \citep{grun:97a}.

\subsection{\ams}

To evaluate whether observations of \ams are in conflict with predictions from rotational disruption theory, we consider below four scenarios through which sub-micron sized dust may exist in the inner solar system.  We find that all four are incompatible with the presence of fast rotational disruption.
 \begin{enumerate}
     \item{Micron-sized dust grains are released farther out in the solar system and drift inwards under the influence of Poynting-Robertson drag.}
     \item{Larger grains are released farther out in the solar system, drift inwards under the influence of Poynting-Robertson drag, and then fragment at 0.1\,au due to collisions or rotational disruption}
    \item{Micron-sized grains are released on bound orbits with pericenter less than 0.1\,au }
     \item{Larger grains are released on bound orbits with pericenter less than 0.1\,au, and fragment due to collisions or rotational disruption.}
 \end{enumerate}

As can be seen from Figure~\ref{GruenPlot}, the rotational disruption time for a 1\,$\mu$m grain is more than four orders of magnitude shorter than the timescale for drift due to Poynting-Robertson drag.  This invalidates Scenario~1 since it is not possible for grains to drift significantly inwards before being disrupted, unless there is a population of grains for which Equation~\eqref{eq:disruptionTime} underestimates the disruption time by several orders of magnitude.

For all grain sizes between 1\,$\mu$m and 1\,cm, Figure~\ref{GruenPlot} shows that the timescale for Poynting-Robertson drag is always at least 100 times longer than the minimum of the two timescales for rotational and collisional disruption.  If rotational disruption were to occur on the timescale given by Equation~\eqref{eq:disruptionTime}, then grains of all relevant sizes would be disrupted in the time it takes them to drift from 1 to 0.1\,au. This invalidates Scenario~2.  However, if orbital averaging greatly reduces the efficiency of the torques for grains greater than a few microns (see Section \ref{sec:discussion}), then this scenario could remain viable.

In considering Scenario~3, we first need to calculate the range of cometary orbits for which such grains would be bound.  If $\beta$ is the ratio of radiation pressure force to gravitational force, then a grain released from a comet on a Keplerian orbit with semi-major axis $a_c$ at orbital radius $r_c$ remains bound if $r_c > 2 \beta a_c$.  $\beta$ depends on the composition and structure of the grain, but for a 1\,$\mu$m grain, it is thought that $\beta \geq 0.2$ \citep{wilck:96a, Silsbee16}.  There are currently (02/09/2025) 32 objects in the JPL small-bodies database\footnote{\url{https://ssd.jpl.nasa.gov/tools/sbdb_query.html}} on bound orbits with pericenter less than 0.1\,au and semi-major axis less than 10\,au.  Of these, all have semi-major axis greater than 0.8\,au.  Therefore, from any of these large bodies, a 1\,$\mu$m dust grain released at pericenter would be gravitationally unbound to the Sun.  

One could imagine the presence of bound micron-sized dust grains at 0.1\,au that were released before or after the cometary pericenter passage.  However, given that the disruption time is a bit shorter than an orbital time, only some fraction of those grains would survive until their next pericenter passage.  In addition, the 32 observed low-pericenter bodies are much fewer than the 22,472 bodies with semi-major axis less than 10\,au and pericenter distance less than 1\,au, or 1,385,016 with semi-major axis less than 10\,au and pericenter distance less than 3\,au, suggesting that several orders of magnitude less dust is produced on orbits with pericenter less than 0.1\,au.  This fact, in combination with the rather fine-tuned scenario needed in order for the dust to remain bound following its release from the comet and survive un-disrupted until pericenter passage interior to 0.1\,au, suggests that direct deposition of grains into orbits with pericenter less than 0.1\,au is responsible for a very small mass of dust.  Further, the observed radial dependence of the zodiacal light is consistent with a dust number density $\propto r^{-1.3}$ in the range from 0.1 to 1\,au \citep[e.g.][]{stenborg:21a, Tsumura23}, which is difficult to reconcile with such a scenario.

Scenario~4 is similar to Scenario~3.  While larger grains would not immediately become unbound, the fraction of them expected to remain on bound orbits following rotational disruption is small.  For example, we simulated a simplified scenario in which $10^6$ dust grains with sizes between 100 and 200\,$\mu$m are released on an orbit with semi-major axis equal to 1\,au, and the eccentricity $e$ uniformly distributed between 0.9 and 1.  The probability of disruption in a particular interval of time is taken to be proportional to the radiative torque at that time (so inversely proportional to the distance from the Sun).  Disrupted grains are assumed to split in two equally sized pieces with initial velocity equal to that of the parent grain.  $\beta$ was assumed to be inversely proportional to grain size, and 0.2 at 1\,$\mu$m.  Successive disruptions took place until the products were less than 1\,$\mu$m in size.  After this exercise, only the products of 178 of the original $10^6$ grains remained on orbits with pericenter less than 0.1\,au.  This indicates that Scenario~4 can only be responsible for a very small amount of dust at 0.1\,au unless there is an unseen population of minor bodies with semi-major axis less than 1\,au.

Having invalidated all four scenarios, we conclude that there is no plausible way to maintain the observed population of \ams if rotational disruption occurs as rapidly as predicted.

\subsection{\bms}
Sub-micron size rotational disruption fragments will likely be gravitationally unbound to the Sun, similar to small collisional fragments. If micron-sized grains are being rotationally disrupted on short timescales, then they should be a significant source of \bms. We demonstrate here that the implied rate of \bmm production far exceeds what has been observed.

To compute the \bmm production rate from rotational disruption, we ran simulations as described in Scenario~4 above, starting with 100-micron sized grains on orbits with eccentricity of 0.5. As grains with radii between 0.1 and 0.4\,$\mu$m have $\beta\geq0.5$ \citep{Silsbee16}, we adopted $\beta =\min(0.2\,\mu{\rm m}/a, 0.5)$.Based on the results in \citet{Silsbee16}, $\beta\geq0.5$ for grains with radii between 0.1 and 0.4\,$\mu$m. We ran simulations as described in scenario~4 above, starting with 100 micron-sized grains on orbits with eccentricity of 0.5, and assuming $\beta = 0.2\,\mu{\rm m}/a.$ We allowed successive rotational disruption events to occur at random values of the grain true anomaly until the size reached below 0.4\,$\mu$m.  We found that 54\% of the fragments ended up on hyperbolic orbits.  We assume that any particle on a hyperbolic orbit is ejected as a $\beta$-meteoroid.  It is also possible that the produced \bms may be further disrupted instead of being ejected from the solar system \citep{Ng25}.  We discuss this possibility in Section \ref{sec:discussion}.  As shown in the previous section, since fast rotational disruption requires that the dust grains be very recently released from comets---and therefore generally on highly eccentric orbits---most rotational disruption cascades would be expected to end with the products gravitationally unbound.

The production rate of \bms in the scenario of fast rotational disruption can be estimated based on the disruption time and the density of bound meteoroids.  As individual dust grains are broken down into smaller grains, we can define the mass flux per unit volume $F_m(m, r)$ as $m n(m, r) \dot m$, where $n(m, r)$ is the differential number density of grains of mass $m$ at location $r$, and $\dot m$ is the expected rate at which a grain loses mass due to centrifugal disruption, here approximated as $m/[2 \tau_{\rm disrupt}(m, r)]$.  Therefore $F_m$ is given by 
\begin{equation}
    F_m = \frac{m^2 n(m, r)}{2 \tau_{\rm disrupt}}\,.
\end{equation}
In a steady-state rotational disruption cascade, $F_m(m, r)$ is independent of $m$.  Using the interplanetary flux model from \citet{Gruen85}, we find that at 1\,au, $F_m$ varies between $5.5 \times 10^{-34}$ and $7.3 \times 10^{-34}$\,g\,s$^{-1}$\,cm$^{-3}$ for masses corresponding to grains in the size range from 1 to 10\,$\mu$m. We take $6 \times 10^{-34}$\,g\,s$^{-1}$\,cm$^{-3}$ as a fiducial value.  Note that $\tau_{\rm disrupt} \propto r^{2}$ provided $\omega_{\rm eq} \gg \omega_{\rm crit}$.  Using the radial density distribution from Equations~(5) and (6) in \citet{stenborg:21a}, we find

\begin{equation}
    F_m = 6 \times 10^{-34} \left(\frac{r}{\rm au}\right)^{-3.3} f\left(\frac{r}{\rm au}\right) {\rm g} \, {\rm s^{-1}} \, {\rm cm^{-3}}
    \label{eq:F_m},
\end{equation}
where 
\begin{equation}
    f(x) = \begin{cases}
1 & \text{if } x \geq 0.088,\\
(x-0.014)/.074  & \text{if } x \in [0.014, 0.088],\\
0  & \text{if } x \leq 0.014.
\end{cases}
\end{equation}
We can relate the flux of \bms at heliocentric radius $R$ to the mass flux interior to $R$:
\begin{equation}
    F_\beta = \int_{0}^R F_m(r) (r/R)^2 dr,
\end{equation}
Using Equation~\eqref{eq:F_m}, we find that 
$F_\beta = 4.6 \times 10^{-20} {\rm g} \, {\rm cm^{-2}} \, {\rm s^{-1}}$.  For comparison to in-situ observations, we convert this mass flux into a number flux assuming a) most conservatively, the characteristic mass of \bms is the largest mass observed by Ulysses of $3 \times 10^{-13}$ g, such that the number flux is $1.5 \times 10^{-7}$ cm$^{-2}$ s$^{-1}$ and b) the characteristic mass is the average mass from the Ulysses mass distribution of $1.7 \times 10^{-14}$ g, with a number flux of $2.8 \times 10^{-6}$ cm$^{-2}$ s$^{-1}$.

\begin{figure*}
\centering
\includegraphics[width = 0.98\textwidth]{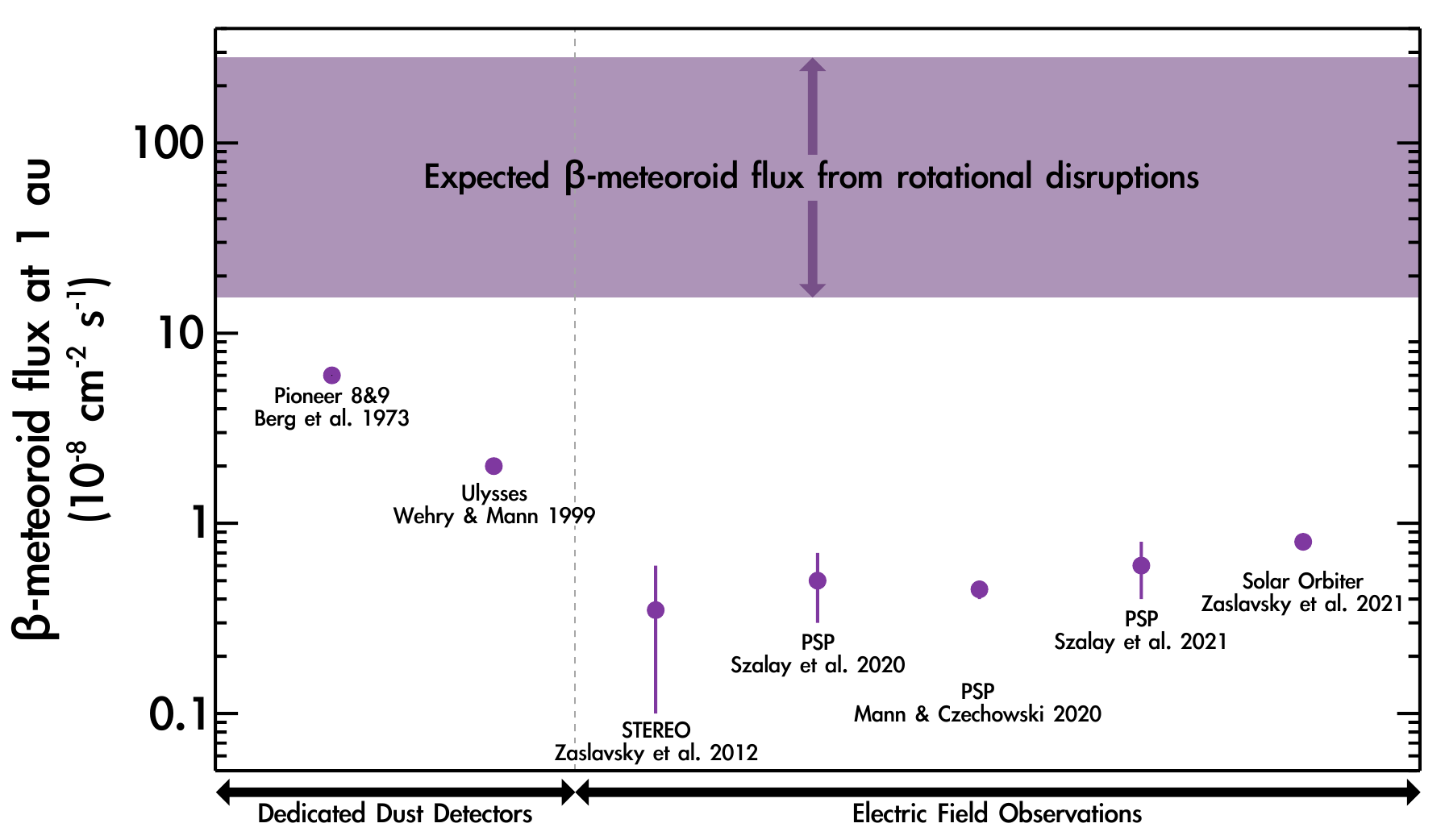}
\caption{\bmm fluxes observed by multiple spacecraft and detection methods \citep{raouafi:23a}. Pioneers~8 \& 9 \citep{berg:73b}, Ulysses \citep{wehry:99a}, as well electric field-based observations from
STEREO \citep{zaslavsky:12a}, SolO \citep{zaslavsky:21a}, and PSP \citep{szalay:20a,szalay:21a,mann:21a}. The purple shaded region shows the estimated \bmm fluxes from  rotational disruption, significantly larger than the observed values.}
 \label{fig:beta_meteoroids}
\end{figure*}

Such a high \bmm production rate is incompatible with in situ measurements. Figure~\ref{fig:beta_meteoroids} shows the fluxes of the unbound, \bmm dust population measured by multiple spacecraft in the inner solar system. All values were converted to reflect the expected $\beta$-meteoroid flux at 1\,au for direct cross-comparison. While there is approximately an order-of-magnitude discrepancy with respect to dedicated dust detectors and electric field observations (left and right of vertical dashed line in Figure~\ref{fig:beta_meteoroids}) the $\beta$-meteoroid fluxes detected close to the sun \citep[PSP;][]{szalay:21a,mann:21a} and at 1\,au \citep[STEREO;][]{zaslavsky:12a} match each other and align well with model $\beta$-meteoroid production rates and fluxes from the collisional grinding of the interplanetary dust cloud near the Sun \citep{Pokorny:2024}. These observed values are all much lower than the predicted \bmm fluxes from rotational disruption. Therefore we find rotational disruption is inconsistent with all existing \bmm observations.

If we assume the number density of meteoroids to be proportional to $\exp{(-2.1 |z/r|)}$, where $z$ is the height above the ecliptic plane \citep{Leinert81}, then the mass flux in the ecliptic plane corresponds to a total mass loss rate at 1\,au of $5.5\times10^7$\,g\,s$^{-1}$. This rate is too large to reconcile with the total mass input that can be supplied by comets.  It was estimated by \citet{Reach07} that the total meteoroid production rate from short-period comets was approximately $3 \times 10^5$\,g\,s$^{-1}$.  It was argued by \citet{Rigley:2022} that the majority of material may be put into the meteoroid population via cometary disintegration events, rather than a slow release in the comets' comas.  Using estimates for the mass contained in the Jupiter family comet population \citep{Tancredi06} and their lifetimes \citep{Levison97, Disisto09}, they found possible production rates of $1.2 - 3.5 \times 10^7$\,g\,s$^{-1}$.  However, these require the entire mass of the disintegrating comet to go into the meteoroid population, whereas a fraction of comets and cometary fragments are presumably lost to planetary collision or ejections from the solar system.

\section{Discussion} \label{sec:discussion}
\begin{figure*}
\centering
\includegraphics[width = \textwidth]{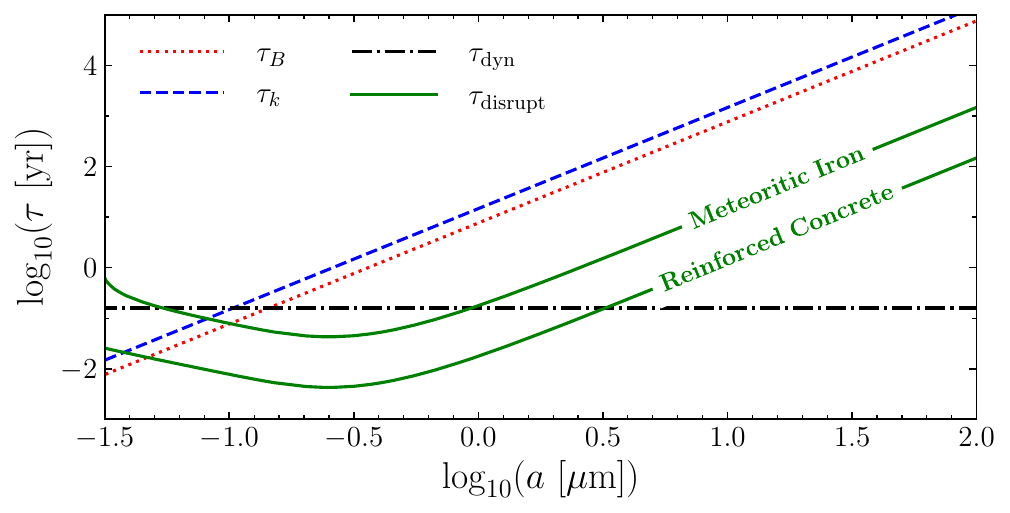}
\caption{Several relevant timescales for grain dynamics.  Red dotted curve: timescale for Larmor precession ($\tau_B$). Blue dashed curve: timescales for precession of grain angular momentum about direction of radiation anisotropy assuming that the grain is spinning at the break-up speed ($\tau_k$; see text).  Black dash-dot curve: dynamical time for grain on a circular orbit at 1\,au (ignoring radiation pressure effects).  Green solid curves: timescale for rotational disruption (see Equation~\eqref{eq:disruptionTime}). The upper curve is for our fiducial $S_{\rm max} = 5\times10^9$\,dyn\,cm$^{-2}$, corresponding to meteoritic iron \citep{Ahles:2021}, while the lower curve is for $S_{\rm max} = 5\times10^7$\,dyn\,cm$^{-2}$, corresponding to reinforced concrete \citep{Arioglu06}.}
 \label{fig:talign}
\end{figure*}

The standard theory for the rotational disruption of grains by radiative torques predicts destruction timescales far too short to be compatible with observations of zodiacal dust. For a wide range of grain sizes, the disruption timescale is much shorter than the orbital time, eliminating effects of orbital averaging as a mitigating factor. While a detailed solution to this problem is beyond the scope of the present study, we suggest that a resolution may be found in consideration of low- and high-$J$ attractor states.

The treatment of radiative torques presented in \citet{Hoang19} is predicated upon the assumption that the grain is driven to an equilibrium orientation such that Equation~\eqref{eq:Qgamma} is a good approximation for the torque.  Such an orientation is called a high-$J$ attractor state \citep{Lazarian:2007}.  Whether a grain has such an attractor depends on its shape, the angle between the radiation and magnetic field, and the grain's magnetic susceptibility \citep{Draine:1996, Lazarian:2007, Lazarian08, Hoang16}.  \citet{Hoang16} showed that nearly 100\% of grains have a high-$J$ attractor point if they contain a significant fraction of metallic iron.   \citet{Herranen21} studied the fraction of an ensemble of model grains with high-$J$ attractors as a function of grain size and the angle between magnetic field and radiation field, finding that the fraction of grains with a high-$J$ attractor state is generally around 50\%, increasing to nearly 100\% for grains with magnetic inclusions. However, the magnetic character of zodiacal dust and, more broadly, the prevalence of low-$J$ attractor states is not well constrained.

For grains having both a high-$J$ and a low-$J$ attractor point, the timescale for a grain to arrive at the high-$J$ attractor is also uncertain.  \citet{Hoang16} argue that grains can be easily knocked out of low-$J$ attractor states via random fluctuations, and therefore all grains that possess a high-$J$ attractor will eventually find themselves in that spin state.  In the context of this work, we caution that ignoring the presence of grains without high-$J$ attractor states, as is frequently done in the literature, likely leads to erroneous conclusions.

One possibility, not explored in this work, is that the grains remaining in the solar system have been selected to possess only low-$J$ attractor points.  If a fraction of  newly produced $\sim \mu$m-sized grains possess a high-$J$ attractor point, then these grains will be rapidly removed from the zodiacal cloud, leaving behind grains possessing only low-$J$ attractor points.  Thus, even if the set of grain shapes produced from comet disintegrations and collisions of larger grains contains a substantial fraction of grains with high-$J$ attractors, the fractional abundance of these grains in the zodiacal cloud will be reduced by several orders of magnitude due to their rapid destruction. Further evolution of the remaining grains will then occur due to collisions and Poynting-Robertson drag, as is commonly assumed.  We note that the principal conclusion of this work---that rotational disruption does not dramatically sculpt dust grain populations---remains unchanged in this scenario.

While the present paper was in the refereeing process, \citet{Ng25} posted a paper considering exactly this model in which only a fraction $f_{\rm high}$ of grains are subject to the centrifugal disruption mechanism.  They compute a steady-state grain size distribution taking into account centrifugal disruption for various values of $f_{\rm high}$. They find that even if $f_{\rm high} = 80\%$, the resulting size distribution is not dramatically different from the case with $f_{\rm high} = 0$, except for an excess of grains with sizes $\leq$20\,nm (see their Figure~5).  This is consistent with our conclusion that unless $f_{\rm high}$ is very close to unity, rotational disruption does not have a significant effect on the size distribution.  We want to emphasize here that $f_{\rm high}$ is the fraction of {\it newly created} grains that possess a high-J attractor point.  The fraction of grains currently present in the solar system with a high-J attractor point will generally be much lower, due to their much faster destruction.

Grains larger than a few microns complete a significant portion of an orbit (or multiple orbits) prior to disruption, and therefore the angle between the principal axis and the radiation field varies.  This should result in some degree of cancellation of the torques over the course of the orbit. Calculating the dynamics in this case is beyond the scope of the present work, but we present here some discussion of the effects of orbit-averaging.

Figure~\ref{fig:talign} presents a comparison of various relevant timescales for grains orbiting at 1\,au. The timescale for Larmor precession is \citep[e.g.,][]{Draine:1997}

\begin{equation}
    \tau_{\rm B} = \frac{4\pi g \mu_{\rm B} \rho a^2}{5\hbar\chi\left(0\right)B}
    \,,
\end{equation}
where $g$ is the gyromagnetic ratio, $\mu_{\rm B}$ is the Bohr magneton, $B$ is the magnetic field strength, and $\chi\left(0\right)$ is the zero-frequency magnetic susceptibility.  Following \citet{Hoang:2014}, We adopt $\chi\left(0\right) = 10^{-4}$ and $B= 50\,\mu$G.

The timescale for precession around the direction of the radiation field anisotropy $\tau_k$ is calculated from Equation~(18) of \citep{Hoang:2014}, assuming $\lambda = 923\,$nm and that $\omega$ is equal to the break-up frequency $\omega_{\rm crit}$ (Equation~\eqref{eq:omega_crit}).  
\par
We conclude that grains smaller than a few microns (with the exact threshold depending on the material strength) experience disruption in much less than an orbital time, and thus the formalism employed in this work adequately describes the rotational dynamics of these grains.

For grains larger than a few microns, the precession times due to the radiation field anisotropy and the magnetic field are long compared both to the orbital time and to the disruption time given in Equation~\eqref{eq:disruptionTime}. This suggests that it is not valid to assume that these grains are aligned with the magnetic field or with the direction of the radiation field. Instead, the grain angular momentum vector remains roughly constant over an orbit. We show in Appendix~\ref{app:orbitAveraging} that if one assumes a constant orientation of the angular momentum vector, the spin-up torque averaged over one orbit is in general non-zero.  However, the conclusion that the precession time is long compared to the spin-up time depends on the particular shape of the grain. In addition, the grain obliquity may also not be fixed.  If the grain is permitted to precess, or the obliquity changes on a timescale comparable to the spin-up timescale, then more complicated dynamics will occur. As our principal conclusions can rest solely on consideration of smaller grains, we do not attempt to model the more complex dynamics of these larger grains in this work.
\par
\citet{Ng25} also raise an interesting point that when rotational disruption is substantially faster than the local dynamical time, grains that are produced in the size range that would normally become \bms will undergo subsequent disruptions prior to being ejected.  This would dramatically enhance the abundance of grains with sizes of 10's of nanometers.  There is no evidence of an enhanced population of grains with radii of 10's of nm in the Ulysses detections (see Figure~\ref{fig:meteoroids}), though the predicted excess is right at the lower edge of its detectable size range. Antenna measurements from the STEREO mission have indicated the possibility of highly time-variable fluxes of nanometer-sized grains picked up in the solar wind \citep{meyer-vernet:09a}, but this population still remains largely unconstrained.  It seems likely that such a dramatic spike in the nano-grain population would drastically reduce the collisional timescale, however modeling the impact of such a decreased collisional lifetime is beyond the scope of this study.  We note that depending on the grain composition, a substantial fraction of even 20\,nm particles may have $\beta$ of several tenths and thus be blown out as $\beta$ meteoroids \citep{Silsbee16}.  The question of whether the fragments are blown out as $\beta$ meteoroids or remain as tiny grains mostly governed by gravity and electromagnetic forces \citep[e.g.,][]{czechowski:10a} does not affect the comparison of the mass flux with the predicted input from cometary bodies.  
\par
A final possibility to reduce the rotational disruption rates is that there is some unknown process taking place during the release of grains from cometary material that leads to their having a nearly spherical shape.  While we cannot rule out this possibility, it is unlikely for a number of reasons.  First, the observed circular polarization of the zodiacal light has been interpreted in the context of non-spherical aligned grains \citep{Wolstencroft:1972, Hoang:2014}.  Second, the grains captured by the Stardust mission from the coma of comet 81P/Wild, as well as micro-meteorites captured in the stratosphere do not look particularly spherical \citep{Brownlee77, Brownlee06}.  Finally, a significant fraction of micron-sized particles likely appear as the product of catastrophic collisions of larger particles.  It is difficult to imagine such a process resulting in nearly spherical particles.

\section{Summary} \label{sec:summary}
The principal conclusions of this work are as follows:

\begin{enumerate}
    \item Using the generally accepted paradigm for rotational disruption presented in \citet{Hoang19}, we calculate the timescale for rotational disruption in the solar system to be less than the dynamical time in the inner solar system for grains between approximately 0.1 and 1\,$\mu$m.  We assume the tensile strength of meteoritic iron, which is almost certainly unrealistically high for actual solar system grains, and therefore our results likely understate the rapidity of the rotational disruption mechanism.
    \item The presence of bound sub-micron-sized dust grains at 0.1\,au is very difficult to explain given the rotational disruption rates unless we posit a substantial portion of the dust population is not subject to the centrifugal disruption mechanism.  Such will be the case even if only a very small fraction of newly created dust grains have this property, since their lifetime will be orders of magnitude longer than those that are subject to centrifugal disruption.
    \item Assuming all micron-sized grains to be subject to rapid rotational disruption, the production rate of sub-micron sized particles is approximately three orders of magnitude above the value estimated from impacts of \bms on the Parker Space Probe, and more than 2 orders of magnitude above the estimated rate at which dust is added to the zodiacal cloud by comets.
\end{enumerate}

These results suggest that the majority of dust grains in the solar system are not subject to the rotational disruption mechanism.  Based on the ratio of the bound \amm dust population to the flux of unbound $\beta$-meteoroids, we conclude that the ratio of bound dust grains that are driven to high-$J$ attractor states must be less than 1\%.  This may be a reflection that high-$J$ attractor states are very rare among realistic grains shapes, or it may represent a selection effect --- grains that had high-$J$ attractors disrupt quickly, leaving behind those that do not.  Further detailed modeling of the solar system dust population including the rotational destruction mechanism is needed to provide more stringent constraints on how it operates.

\section*{Acknowledgments}
We thank B. Draine for discussions that inspired this study and the anonymous referee for comments that improved the manuscript. J. R. Szalay acknowledges NASA Parker Solar Probe Guest Investigator grant 80NSSC21K1764. This research was carried out in part at the Jet Propulsion Laboratory, California Institute of Technology, under a contract with the National Aeronautics and Space Administration (80NM0018D0004). J.-G.K is supported by KIAS Individual Grant QP098701 at Korea Institute for Advanced Study. P. P. acknowledges support provided by NASA’s Planetary Science Division Research Program, through ISFM work packages EIMM and Planetary Geodesy at NASA Goddard Space Flight Center, NASA award numbers 80GSFC24M0006 and 80NSSC21K0153. 

\software{Matplotlib \citep{Matplotlib}, Matplotlib label lines \citep{Cadiou:2022}, NumPy \citep{NumPy, numpy:2020}, SciPy \citep{SciPy}}

\facilities{Parker}

\bibliographystyle{apj}
\bibliography{solarSystem}

\counterwithin*{equation}{section}
\renewcommand\theequation{\thesection\arabic{equation}}
\appendix

\section{orbit-averaging of torques}
\label{app:orbitAveraging}
In this appendix, we show that one would not in general expect the torque to average to zero over an orbit, provided that the direction of the angular momentum vector remains constant over the orbit.  To analyze this situation, we consider the analytical model presented in \citet{Lazarian:2007} consisting of a spheroidal grain with a mirror attached by a pole. The instantaneous torque on the grain is given by their Equations~B.14 and B.18--B.20 as
\begin{equation}
    \Gamma_{\rm rad} =  2\gamma \mu_{\rm rad} l_2^2 l_1 {\bf Q}
\end{equation}
where $l_1$ and $l_2$ are constants describing the mirror and ${\bf Q}$ has components 
\begin{widetext}
\begin{align}
   Q_x &= \frac{1}{2}|\sin{\alpha} \cos{\theta} - \cos{\alpha} \sin{\theta} \cos{\beta}| \left[\sin{(2\alpha)} \cos^2{\theta} - \cos{2 \alpha} \cos{\beta} \sin{2 \theta} - \sin{(2 \alpha)} \sin^2{(\theta)} \cos^2{(\beta)}\right] \nonumber\\
   Q_y &= -|\sin{\alpha} \cos{\theta} - \cos{\alpha} \sin{\theta} \cos{\beta}| [\sin^2{(\alpha)}\cos{(\beta)} \cos^2{(\theta)} - \nonumber \\
   & \quad \quad \frac{1}{4} \sin{(2\alpha)} [1+\cos^2{(\beta)}] \sin{(2 \theta)} + \cos^2{(\alpha)} \cos{(\beta)} \sin^2{(\theta)}] \nonumber\\
   Q_z &= -|\sin{\alpha} \cos{\theta} - \cos{\alpha} \sin{\theta} \cos{\beta}| \sin(\alpha) \sin(\beta) \left[\sin{\alpha}\cos{\theta} - \cos{\alpha} \cos{\beta} \sin{\theta}\right]
   \,.
   \label{eq:AMOTorqueComponents}
\end{align}
\end{widetext}
Here $\theta$ is the angle between the direction of the radiation field anisotropy and the principal axis of the grain, $\beta$ is the angle of the grain rotation about the principal axis, and $\alpha$ is an angle describing the orientation of the mirror. If the pole points in the direction of $\hat a_3$, then the normal to the mirror has component $\sin{\alpha}$ in the direction of $\hat a_1$ and $\cos{\alpha}$ in the direction of $\hat a_2$, where $\hat a_1$ is the direction of the principal moment of inertia, and $\hat a_2$ and $\hat a_3$ are two perpendicular unit vectors.  The $\hat x$ direction corresponds to the direction of radiation field anisotropy, and the principal axis $a_1$ lies in the x-y plane.

\citet{Lazarian:2007} define the spin-up torque as the component of the torque along the principal axis (see their Equation~(32)):
\begin{equation}
    \Gamma_{\rm spinup} = \Gamma_{\rm rad}^x \cos{\theta} + \Gamma_{\rm rad}^y \sin{\theta}.
    \label{eq: spin-up torque}
\end{equation}

Imagine a system in which the principal axis of the grain is fixed in 3D space, tilted at angle $\psi$ with respect to the orbital plane.  Let $\chi$ be the azimuthal angle describing the grain's orbit, with $\chi = 0$ defined such that $\theta = \psi$ at $\chi = 0$ (i.e., at $\chi = 0$, $a_1$ has no azimuthal component).  Then we may write 
\begin{equation}
    \theta(\chi) = \cos^{-1}{(\cos{\chi} \cos{\psi})}.
\end{equation}
We can then calculate the orbit-averaged torque as 
\begin{equation}
\langle \Gamma_{\rm spinup} \rangle = \left(\frac{1}{2\pi}\right)^2 \int_{\beta = 0}^{2\pi} \int_{\chi = 0}^{2 \pi} \Gamma_{\rm spinup} (\beta, \theta(\chi)) d \beta d \chi\,.
\end{equation}
Numerically, we find this integral to be zero for all values of $\alpha$ and $\psi$.  Introducing an orbital eccentricity would not help: since both angular velocity and the radiation field are inversely proportional to the squared orbital radius, the effect of increased torque near pericenter is exactly cancelled by the reduced time spent near pericenter, and the orbit-averaged spin-up torque remains zero.

We could imagine also an analytic model in which the attached mirror does not reflect on both sides, but instead absorbs photons which are incident on one side, and reflects photons incident on the other.  We may assume the mirror to emit infrared photons equally from both sides so that emission of photons from the mirror does not contribute to the net torque.  The radiation pressure due to light incident on the absorbing side of such a mirror is 1/2 that due to light incident on the reflecting side.  In the context of the equations in \citet{Lazarian:2007}, this means that the term $|\hat e_1 \cdot \hat N|$ in their Equation~B.10 should be replaced with $\mathscr{F}(\hat e_1 \cdot \hat N)$, where $\mathscr{F}(x) \equiv |x|[1 + \mathcal{H}(x)]/2$, where $\mathcal{H}(x)$ is the Heaviside function.  Carrying that change over to Equations~\eqref{eq:AMOTorqueComponents} of this work, each prefactor $|\sin{\alpha} \cos{\theta} - \cos{\alpha} \sin{\theta} \cos{\beta}|$ must be replaced by $\mathscr{F}(\sin{\alpha} \cos{\theta} - \cos{\alpha} \sin{\theta} \cos{\beta})$.  Making that change, the torque in the analytic model no longer vanishes when averaged over $\chi$.  In fact, for the simplest case in which $\psi = 0$, we can calculate 
\begin{equation}
   \mathscr{R}(\alpha) = \frac{\int_{\theta = 0}^{2\pi} \int_{\beta = 0}^{2\pi} \Gamma_{\rm spinup}(\alpha, \beta, \theta)}{\int_{\theta = 0}^{2\pi} |\int_{\beta = 0}^{2\pi} \Gamma_{\rm spinup}(\alpha, \beta, \theta)|}.
\end{equation}
Numerically we find that $\mathscr{R}(\alpha)$ varies between 0.33 and 1, depending on the value of $\alpha$, suggesting that for such a toy grain model, indeed the torques are lowered by averaging over an orbit, but they are by no means eliminated.
\par
We caution that in this analysis we have assumed that the spin-up timescale is short compared with the timescale for the orientation of the principal axis of the grain to change.  It was argued in \citet{Rubincam:2000} that this approximation is not valid for asteroids, and in fact, they may undergo cycles of changing angular momentum and obliquity, termed ``YORP cycles.''  Investigating whether a similar effect holds for dust grains is an intriguing possibility, however is beyond the scope of this work.  For this reason, we limit our principal conclusions to grains below a few microns in size, where the predicted spin-up times are short compared with an orbital time.

\end{document}